\documentclass[sigconf]{acmart}

\usepackage{url}
\usepackage{soul}
\usepackage{xcolor}

\definecolor{ao(english)}{rgb}{0.0, 0, 1}

\newcommand{\chiadd}[1]{\textcolor{black}{#1}}

\newcommand{\chiremove}[1]{}

\AtBeginDocument{%
  \providecommand\BibTeX{{%
    \normalfont B\kern-0.5em{\scshape i\kern-0.25em b}\kern-0.8em\TeX}}}
\AtBeginEnvironment{quote}{\itshape}

\copyrightyear{2023}
\acmYear{2023}
\setcopyright{acmlicensed}\acmConference[CHI '23]{Proceedings of the 2023 CHI Conference on Human Factors in Computing Systems}{April 23--28, 2023}{Hamburg, Germany}
\acmBooktitle{Proceedings of the 2023 CHI Conference on Human Factors in Computing Systems (CHI '23), April 23--28, 2023, Hamburg, Germany}
\acmPrice{15.00}
\acmDOI{10.1145/3544548.3581542}
\acmISBN{978-1-4503-9421-5/23/04}

\begin{document}

\title{\chiremove{Gauge against the Machine: A Study of} \chiadd{Understanding} Journalists' Workflows in News Curation}




\author{Shubham Atreja}
\authornote{This research began during the author's internship at Microsoft Research, Bangalore, India}
\affiliation{
 \institution{University of Michigan School of Information}
 \streetaddress{105 S. State St.}
 \city{Ann Arbor}
 \state{Michigan}
 \country{USA}
 \postcode{48103}
}
\email{satreja@umich.edu}

\author{Shruthi Srinath}
\affiliation{
 \institution{Microsoft Research}
 \city{Bangalore}
 \country{India}
}
\email{shruthisrinath27@gmail.com}

\author{Mohit Jain}
\affiliation{
 \institution{Microsoft Research}
 \city{Bangalore}
 \country{India}
}
\email{mohja@microsoft.com}

\author{Joyojeet Pal}
\affiliation{
 \institution{University of Michigan School of Information}
 \streetaddress{105 S. State St.}
 \city{Ann Arbor}
 \state{Michigan}
 \country{USA}
 \postcode{48103}
}
\email{joyojeet@umich.edu}



\renewcommand{\shortauthors}{Atreja, et al.}

\begin{abstract}
    
With the increasing dominance of internet as a source of news consumption, there has been a rise in the production and popularity of email newsletters compiled by individual journalists. However, there is little research on the processes of aggregation, and how these differ between expert journalists and trained machines. \chiadd{In this paper,} we interviewed journalists who curate newsletters from around the world. \chiadd{Through an in-depth understanding of journalists' workflows, our findings lay out the role of their prior experience in the value they bring into the curation process, their own use of algorithms in finding stories for their newsletter, and their internalization of their readers' interests and the context they are curating for.}\chiremove{ and found that journalists seek to balance the consistency of repeating popular content with human values of affect. We provide design insights on how technology can support the work of these experts.} While identifying the role of human expertise, we highlight the importance of hybrid curation and provide design insights on how technology can support the work of these experts.

\end{abstract}

\ccsdesc[300]{Human-centered computing~User studies}



\maketitle

\section{Introduction}

The Digital News Report 2022 by the Reuters Institute notes that the internet has become a major source of news consumption the world over~\cite{Newman_Fletcher_Robertson_Eddy_Nielsen_2022}. In the United States, it's the largest source of news consumption, ahead of TV and print media \cite{Newman_Fletcher_Robertson_Eddy_Nielsen_2022}. While online news consumption is on the rise, direct access to news websites continues to decline, and intermediaries that rely on algorithmic aggregation -- such as Google News, Bing News, and Apple News -- are becoming the mainstay of daily news consumption \cite{statistaLeadingNews, Newman_Fletcher_Robertson_Eddy_Nielsen_2022}. 


While this rise in digital news has been accompanied by a precipitous fall in printed news from local or regional sources \cite{nytimesOverNewspapers}, a new form of curated news has emerged -- independent newsletters. These newsletters are delivered over email and typically put together by professional journalists who source their news from multiple publishers. The growth and hyper-customizability of newsletters have in part been driven by publishing platforms. Substack, the leading platform, has garnered the attention of over 10 million users, with 1 million paid subscribers \cite{fortuneSubstackersMaking}. The growth of newsletters provides new hopes and concerns for the frontiers of independent, human-driven journalism \cite{shrivastava2022role}.

\chiremove{These emerging patterns raise important questions about the functioning of both journalists and algorithms as aggregators of news. }\chiadd{Like algorithms, journalist curated newsletters are another form of intermediary impacting the visibility and consumption of published news. This gatekeeping function \cite{wallace2018modelling, thorson2016curated} raises important questions about the practices of individual news curators and algorithmic aggregators as gatekeepers of already published news.} How do they get their content? What factors guide the selection of their content? And how do humans and algorithms differ on these aspects? To that end, algorithmic aggregators have already received significant attention from researchers and practitioners alike \cite{Bandy_Diakopoulos_2020,Bandy_Diakopoulos_2021, devito2017editors, nikolov2019quantifying,nechushtai2019kind,trielli2019search}. The successful and widespread adaption of algorithmic aggregators like Google News underlines the extent of implicit knowledge and research that is put into their design. In addition, algorithmic audits have been conducted by researchers as a means to analyze the outputs of algorithmic aggregation \cite{trielli2019search,Bandy_Diakopoulos_2020}.

Meanwhile, newsletters curated by journalists have received little attention from the research community. In this study, we provide a detailed exploration of journalists' news curation practices. Through in-depth interviews, we investigate the various steps of their workflow, and their thought processes that define the selection and presentation of news. Our findings lay out the role of journalists' prior experience in the value they bring into the reading process, \chiremove{their own use of algorithms in finding stories for their newsletter, }and their internalization of subjectivity and readability as it relates to their choices as well as their readers' interests. 

By contrasting these findings with the workings of algorithms, the paper presents a meditation into the distinctions between the ways machines understand reader predispositions with how a human chooses for another. We examine how human curation is not entirely human -- in that algorithms are a necessary part of the hybrid and evolving workflows behind journalists' news curation, just as algorithmic processes are dependent on the human in the loop.\chiremove{ In conclusion, we propose a set of design implications for tools that can support and supplant the journalists' existing practices of news curation.}

\chiadd{
In summary, this paper makes the following contributions:
\begin{itemize}
    \item We enumerate how journalists curate their stories by navigating the ecosystem of online news which includes both individual content producers and intermediaries like algorithmic aggregators and social media platforms.
    \item We list the factors that guide the journalists’ curation process, the reliance on prior journalistic experience, their individual sense of the readers’ interests, and finally, the context they are curating for. 
    \item We highlight the nuances of hybrid curation — involving both algorithms and humans — and identify ways in which algorithms can support the journalists’ existing practices of news curation.
\end{itemize}
}

\section{Background and Related Work}

With the abundance of news available online, navigating one's news consumption has become non-trivial. Thus, a significant portion of online news consumption is driven by intermediaries, which include algorithmic aggregators like Google News, and Apple News, social media platforms like Facebook and Twitter, and more recently, newsletters curated by journalists \cite{hendrickx2020innovating,Newman_Fletcher_Robertson_Eddy_Nielsen_2022}. In Section~\ref{sec:newsletters}, we provide a background on newsletters curated by journalists, which are the subject of this study. Section~\ref{sec:algos} is devoted to algorithmic aggregators as they offer a direct contrast to the work done by journalists. Finally, in Section~\ref{sec:hci}, we provide a roundup of recent research on the practices of journalism within the field of HCI. 

\subsection{Study Context: Manually Curated Newsletters}
\label{sec:newsletters}
Though the concept of newsletter can be traced back to traditional journalistic practices, today's newsletters---delivered over email---offers a variety of unique advantages. For journalists who may work individually or in teams to curate their newsletters, the key benefits are monetary income with editorial freedom \cite{shrivastava2022role}. Platforms like Substack and Revue enable journalists to create and monetize their newsletters through a subscription-based model. Journalists, many of whom have left news organizations to start their own newsletter \cite{washingtonpostReportersLeaving}, particularly enjoy the editorial freedom of selecting their own stories, providing their perspectives, adopting new writing styles, and more importantly, developing their author identity \cite{washingtonpostReportersLeaving,niemanlabFromNewsroom}.

In addition, these factors translate into direct benefits for the readers as well. The Digital News Report by the Reuters Institutes uncovered that, apart from the convenience of the email format, readers enjoy the diverse perspectives and unique content offered by the newsletters, as well as the personality and voice of the curator \cite{Newman_Fletcher_Robertson_Eddy_Nielsen_2022}. Therefore, as prior work has noted, newsletters curated by journalists are more than just a compilation of stories \cite{nytlicensingWhatNews,seely2021email}, and often involve ``representing and reformulating'' \cite{howarth2015exploring} existing news stories in new configurations \cite{cui2017does}. That is unlike algorithmic aggregation, which is limited to providing a list of news stories for their readers.

The growing popularity of newsletters is further underlined by the emergence of digital news organizations specializing in curating newsletters on specific topics, such as theSkimm\footnote{https://www.theskimm.com/} and Axios\footnote{https://www.axios.com/newsletters}. Even mainstream news organisations recognize the potential of newsletters, and hence are investing resources to produce email-based newsletters.
The New York Times, for example, now produces more than 50 different email newsletters every week on a variety of topics which are read by $\sim$15 million people \cite{niemanlabYorkTimes}.

\subsection{Algorithmic News Aggregation}
\label{sec:algos}

\chiremove{Given }The increasingly important role of algorithmic aggregators in news consumption \chiadd{has led to a consideration of algorithmic influences in recent gatekeeping models, including how they access information and their selection criteria \cite{wallace2018modelling, thorson2016curated, lewis2015actors}. Within the specific context of aggregation,} we discuss \chiadd{how these algorithms get access to news stories and} factors that drive them to select and rank these stories. The description is knowingly incomplete, as these intermediaries reveal minimal details about their algorithms, usually buried across multiple pages in their help section. In addition, we rely on 
empirical findings from prior work that analyzed news stories picked by Google News \cite{trielli2019search, nechushtai2019kind, nikolov2019quantifying,haim2018burst}\footnote{\chiadd{Out of the four cited studies, two considered Google News recommendations \cite{nechushtai2019kind,haim2018burst}, one considered Google search results \cite{trielli2019search}, while the last one considered both \cite{nikolov2019quantifying}. While these are two different features provided by Google, most of our participants did not draw a distinction between them, and a qualitative understanding of the findings from these studies also shows no differences. Therefore, we consider them the same for the rest of our analysis.}} and Apple News \cite{Bandy_Diakopoulos_2020}.


\textbf{News Sources}: Algorithmic aggregators use automated crawlers to collect news from thousands of content producers, e.g., 
Google aggregates news from over 20,000 content producers \cite{searchenginejournalAvoidThese}. These aggregator platforms have specified a set of guidelines to check for the eligibility of news content.
For instance, Google News ignores dangerous, hateful, and sexually explicit content, while prefers news stories with transparent data around author name, published date, and publisher
\cite{contentpolicy}. However, satisfying the criteria does not guarantee that the news content will be ranked; the decision often lies with the aggregator and content producers can do little to influence that decision
\cite{noGuarantee}.



\textbf{News Selection and Ranking}:
After collecting hundreds of thousands of stories, the top few are shown to the end users in a ranked list. There are multiple factors influencing the selection and ranking of news stories, including prominence, authoritativeness, and freshness of the story.


Prominence is computed by the amount of coverage a story is receiving from different content producers, and also the placement of the story on the content producers' website
\cite{newsinitiativeSurfacingUseful}. Inadvertently, it can lead to a popularity bias \cite{nikolov2019quantifying} as stories about major events and/or stories that are already popular 
are likely to be ranked at the top.
\citet{Bandy_Diakopoulos_2020} performed an algorithmic audit of Apple News' `Trending Stories' in the US and found that most stories featured prominent celebrities (like Justin Bieber, Kate Middleton) or politicians (like Alexandria Ocasio-Cortez, Donald Trump), whereas stories aggregated by human journalists during the same period were more likely to feature policy issues and international events.

Authoritativeness of the content is crucial to select stories from reliable sources. Both Google News and Bing News rely on a variety of signals to check for authoritativeness, such as the reputation of the author, publisher, and people quoted in the story. Google, in particular, hires thousands of human raters \cite{blogOverviewRater} to evaluate news stories according to its Expertise-Authoritativeness-Trustworthiness (EAT) framework \cite{eat}. 
These raters are not hired for their journalistic expertise but to act as independent sensors trained to evaluate the content according to the criteria defined by Google. Empirical findings \cite{trielli2019search,nikolov2019quantifying,nechushtai2019kind} suggest that algorithmic aggregation disproportionately favors content from a few mainstream content producers. For instance, \citet{nechushtai2019kind} found that 69\% of top news stories on Google News were from five news producers -- all of whom were national organizations. Apple News' `Trending Stories' exhibit a similar bias with three content producers contributing 45\% of the news stories~\cite{Bandy_Diakopoulos_2020}. 


While ranking stories, algorithms also consider the freshness of the content, favoring recent content in a bid to provide up-to-date information \cite{bingBingWebmaster,newsinitiativeSurfacingUseful}. 
An analysis of Google News found that 83\% of the news stories were less than 24 hours old, while 13\% were less than 1 hour old \cite{trielli2019search}. Favoring recent content can result in a high churn of news stories. Apple News' `Trending Stories' section on an average displayed $\sim$50 stories in a day, compared to human aggregators hired by Apple who only shared $\sim$20 stories a day \cite{Bandy_Diakopoulos_2020}. Prior research \cite{chakraborty2015can} found that a high churn of stories biased news selection towards specific topics (mainly sports and politics) that received regular updates, while topics like environment, which are not covered as broadly, received less coverage.

\textbf{News Personalization}: Prior research has found little evidence of personalization in the top stories ranked by both Google News \cite{nechushtai2019kind,haim2018burst} and Apple News \cite{Bandy_Diakopoulos_2020} as the set of stories seen by different users were largely similar. As noted on the Google News website, however, the personalization in terms of users' interests may influence the set of stories for which they receive a push notification. This is unlike newsletters curated by journalists, where all readers are sent the same set of stories. In the absence of empirical evidence on push notifications sent by news aggregators, it is unclear if such personalization has the potential to introduce additional biases in news recommendation. 

Overall, various biases --- popularity bias \cite{nikolov2019quantifying}, homogeneity bias \cite{nikolov2019quantifying}, and coverage bias \cite{chakraborty2015can} --- have been uncovered in algorithmic aggregation by prior work. These biases could be a reflection of biases that exist among humans (e.g., popularity bias), or a reflection of the algorithm's encoded logic of what is relevant knowledge \cite{gillespie2014relevance} (e.g., homogeneity bias). Given the lack of objectivity in algorithmic news aggregation, it becomes crucial to understand the journalists' news curation process, with an emphasis on the role of their 
subjective judgment in this process. The analysis of algorithmic aggregation also suggests that their logic undermines the importance of journalistic values, such as diversity and novelty \cite{Bandy_Diakopoulos_2021}. Through our work, we also contribute detailed evidence on the practices adopted by journalists to uphold these values. 


\subsection{Journalism and HCI}
\label{sec:hci}

\chiadd{We have seen a growing body of work on journalism in HCI \cite{aitamurto2019hci, cohen2011computational, Diakopoulos_Trielli_Lee_2021, Oh_Choi_Lee_Park_Kim_Song_Kim_Lee_Suh_2020, Wang_Diakopoulos_2021, Smith_Wang_Karumur_Zhu_2018, Tolmie_Procter_2017} at a time when online aggregation and algorithmic feeds of news stories have dramatically changed both the news consumption environment and the professional practice of mainstream journalism \cite{cohen2011computational,diakopoulos2019automating}. The online information environment, characterized by a rapid access to updates from across the world, disproportionately favors the ``breaking news'' form of viral information, often driven by online engagement \cite{hosni2020minimizing}. Furthermore, the role of networks in popularizing news items has made recommender systems and the public reaction to them central elements of what drives newsworthiness rather than journalists' editorial choices \cite{kumpel2015news, o2021twitter}. These developments have fundamentally changed the writing and editing processes in mainstream journalism, creating an ever-growing interdependence between the journalists seeking to draw attention to their content, and the algorithms driving that content \cite{molyneux2022legitimating}.}

\chiadd{The contemporary news cycle, with a regular access to live updates from around the world, increases the pressure on journalists and media houses to constantly keep track of events, both adding new content multiple times a day, and editing existing content as it changes.  However, while the internet affords access to a much larger pool of information, it can be challenging for journalists to make sense of all this information and to separate noise from useful information. }

\chiadd{Consequently, tools have been developed to support various stages of a journalists’ work — from news discovery \cite{Diakopoulos_Trielli_Lee_2021} to news production \cite{Oh_Choi_Lee_Park_Kim_Song_Kim_Lee_Suh_2020} and delivery \cite{ Bentley_Quehl_Wirfs-Brock_Bica_2019,Flintham_Karner_Bachour_Creswick_Gupta_Moran_2018}. For example, \citet{Diakopoulos_Trielli_Lee_2021} developed Algorithm Tips, a tool designed to support news discovery by helping journalists find newsworthy leads on algorithmic decision-making systems being used across all levels of the US government. \citet{Oh_Choi_Lee_Park_Kim_Song_Kim_Lee_Suh_2020} developed NewsRobot to automatically generate news stories at a scale as major events are unfolding in real time and \citet{Wang_Diakopoulos_2021} developed a tool to analyze large quantities of user-generated content to support journalists' discovery of news sources from their audience.}

\chiadd{In this study, we focus on the relatively new expansion of online newsletters, that have grown massively with the entry of independent publishing platforms like Substack into the armory of journalists \cite{hobbs2021substack}. Journalists who curate their news from a wide range of publishers are not subject to the constraints of any one news organization and have access to a much large set of news stories. However, the large universe of news stories published everyday also makes their task challenging and creates avenues for technological (or algorithmic) support. }

\chiadd{When designing new technologies to fulfill journalistic goals, both journalism and HCI scholars have argued for the need to develop a deeper understanding of the journalists’ sociotechnical contexts \cite{Smith_Wang_Karumur_Zhu_2018,Diakopoulos_2020, Tolmie_Procter_2017, bucher2017machines}. This line of work has further emphasized the importance of embedding journalistic values as part of new technologies and acknowledging the importance of journalistic judgment in their workflows \cite{Diakopoulos_2019, carlson2018automating}. For instance, \citet{bucher2017machines} presented a rich description of journalists and algorithms working together to update the homepage of a news app. Even though algorithms made the final call on stories that were displayed to a user, the journalists' subjective judgments about relevance and newsworthiness of a story were important features that influenced the algorithmic output. While scholars have underlined the importance of these hybrid workflows between journalists and algorithms, challenges still exist on how to integrate algorithmic outputs in journalisic workflows. Our work contributes to this body of work, as we provide an in-depth understanding of the journalists' existing news curation practices, and informed by this understanding, explore the design space of solutions to support the journalists’ task of news curation. }

\section{Methods}
\begin{table*}[t]
\caption{Demographic details of our participants.
}
\resizebox{1.0\textwidth}{!}{
\begin{tabular}{p{0.04\textwidth} p{0.06\textwidth} p{0.07\textwidth} p{0.09\textwidth} p{0.3\textwidth} p{0.15\textwidth} p{0.07\textwidth} p{0.09\textwidth} p{0.1\textwidth} }
\hline
\textbf{Code} & \textbf{Gender} & \textbf{Country}  & \textbf{Experience} & \textbf{News Category} & \textbf{Geographical Focus} & \textbf{No. of Stories} & \textbf{Subscriber Base} & \textbf{Newsletter Frequency} \\ \hline
P1   & F      & India & 3 years                  & All                              & Non-western          & 1   & 200              & Biweekly             \\ 
P1   & F      & India & 3 years                  & Data                                & India               & 5-7    & 14,000              & Biweekly             \\ 
P2   & M      & US       & 8 years                  & Data                                & Global                & 4-7   & 50,000      & Daily                \\ 
P3   & F      & India    & 4 years                  & All                             & India / Global         & 5-7    & 5,000              & Daily                \\ 
P4   & M      & India    & 2 years                  & Business, Technology, and Policy    & India / Global         & 10-12  & 25,000             & Daily                \\ 
P5   & M      & UK       & 12 years                 & All                        & Global               & 5     & 90,000             & Daily                \\ 
P6   & M      & India    & 2 years                  & Business, Technology, and Policy    & India / Global         & 10-12 & 25,000           & Daily                \\ 
P7   & M      & US       & 6 years                  & Finance, International, and Science & US / Global & 15-20  & 12,000            & Daily                \\ 
P8   & M      & US       & 8 years                  & All                                 & Global               & 10-12  & --              & Daily                \\ 
P8   & M      & US       & 8 years                  & Local                               & A city in the US        & 4-6             & 20,000             & Daily                \\ 
P9   & F      & Italy    & 4 months               & All                                 & North America        & 6-8    & 500,000           & Daily                \\ 
P10   & F      & US    & 2 years               & Technology and Consumer Culture                                 & Global        & 5-10    & 50,000          & Weekly                \\ \hline
\end{tabular}
}
\label{tab:participants}

\end{table*}

For the study, our sample was restricted to journalists who curate newsletters on a regular basis by combining stories from many news publishers. \chiadd{Journalists who curate stories from different publishers have the freedom to select stories from venues of their choosing, and are free of any organizational constraints of one publication, such as the history, the readership, and the pagewise limitation of that publication. By extension, curation for a single publication is constrained to what that organization can produce or syndicate. On the other hand, the journalists we interviewed can provide a more diverse perspective to their readers. This in turn means that their universe for selection is much broader, and thus exhibits greater variance in source type, quality, and readership, making it a challenging use case and requiring a structured curation process.}
However, this also leaves a relatively small universe of journalists, and understanding their professional practices requires an in-depth study of their daily work. Therefore, we conducted in-depth semi-structured interviews with 10 participants who were collectively involved in curating 11 newsletters. 
The main focus of the interviews was to understand their workflow (practices), their process of selecting stories, and their reflective practice of understanding their readership. All our interviewees had been in journalism for several years. Our study was approved by IRB.

\subsection{Participants}
We recruited participants based on our direct knowledge of such newsletters, by browsing through Substack\footnote{https://substack.com/} (a platform that allows independent writes to run their own newsletters), and, a combination of snowball, and purposive sampling. We ensured that our participant pool provided us with a diverse experience by considering newsletters that covered different contexts, in terms of their geographical focus as well as the news topics that they covered. 

To be included in our sample, participants needed to fulfill the following conditions. First, they needed professional experience in journalism, second, they needed to be directly involved with aggregating stories for a newsletter, and third, the newsletter should not be restricted to a particular news publisher. Thus individuals who performed editorial work within a publication, or aggregated stories through newswires were excluded from recruitment.

Table \ref{tab:participants} shows the participant details. We interviewed 10 participants who cover 11 different newsletters. Two participants (P1 and P8) have managed two newsletters each, while two other participants (P4 and P6) were associated with the same newsletter. Participants came from different countries (India, UK, US, Italy) and varied in terms of their journalistic experience (2 years to 27 years), the popularity of their newsletter (200 to 500,000 subscribers),  as well as their experience with running a newsletter (4 months to 12 years). One participant (P8) did not reveal the size of the subscriber base for one of their newsletters. All participants were offered a compensation of USD \$40 gift card (or an equivalent amount in their local currency). Two participants respectfully denied the compensation. The names of the newsletters are not revealed to protect the participants' anonymity. 

\subsection{Interview Protocol}

The interviews were conducted over Zoom between June and August, 2022. Each interview was about an hour long. All interviews were \chiadd{conducted in English and} were audio recorded with the permission of the participants.

We started the interview by asking participants to describe their process for curating the newsletter in a step-by-step manner. As the participants described their process, we asked follow-up questions about their use of technological tools and other resources during the process. Next, we asked participants to list down some of the factors they consider when aggregating stories from different sources. For this part of the interview, we centered the narrative around previous editions of their newsletters, asking questions about why or how particular stories were selected. At the end of the interview, participants were asked to identify what according to them was the most difficult part of the process, and the most important part of the process.

\subsection{Data analysis}
\chiremove{We first transcribed the audio interviews}\chiadd{The first author conducted all the interviews for this study, and transcribed them soon after they were conducted.}
\chiadd{To systematically analyze the interview data, we used grounded theory, as outlined by Glaser~\cite{glaser-grounded}.}
\chiadd{We subjected our data to open coding in}
\chiremove{ and followed}an inductive and interpretive \chiremove{coding approach}\chiadd{manner}~\cite{merriam2019qualitative}\chiremove{ to this data.}\chiadd{, and rigorously categorized our codes to examine the workflow and practices of newsletter creators.}
Three authors\chiremove{periodically discussed the interview data to identify emerging themes.}\chiadd{, co-located in the same working space, regularly participated in the coding process and iterated upon the codes until consensus was reached. Over the course of analysis, they met over multiple days to: (1) discuss coding plans, (2) develop preliminary codebook, (3) review the codebook and refine/edit codes, and (4) finalize categories and themes.}
\chiremove{Initial themes included}\chiadd{The first-level codes were specific, such as} ``balancing the composition'', ``selecting news frames'', and ``social media''.
\chiremove{We iterated over this data to produce}\chiadd{After several rounds of iteration, the codes were condensed into}
high-level themes, such as ``story aggregation'' and ``value-addition''.
\chiadd{Based on that, we}\chiremove{Ultimately, we decided to} structured the findings around how journalists navigate the news ecosystem, and the factors influencing their decisions and actions\chiremove{ The factors abstracted out to three main themes:}\chiadd{, mainly} journalistic experience, readers' interests, and subjective judgments.
\chiadd{Several other approaches including, multi-level code reviews, peer debriefing, and member checks, were used to improve the credibility and validity of the coding process. Appendix \ref{appendix} shows our final codebook.}

\chiremove{When describing}\chiadd{Note: In} our findings, we use the term ``journalists'' to imply journalists involved in the curation of newsletters.

\subsection{Positionality}
All authors are of Indian origin. Three authors identify as male and one as female. Two authors have been journalists; one has worked at a wire desk selecting stories for a mainstream newspaper and the other has worked as a copy-editor and reporter for a  newspaper. 
Two other authors have prior experience with human-centered approaches for designing new technologies. 
The research we present in this work was motivated by a larger project around promoting news consumption at an organization level. Having worked in journalism, the authors were drawn to investigate how journalists curate newsletters, and what would it mean to curate newsletters at an organizational level. Our long-term goal is to design new tools that can support journalists' news curation workflows.

\subsection{Limitations}
Our findings are derived from a relatively small sample of 10 journalists. However, to understand the challenges of handling a large volume of news stories, we intentionally excluded journalists whose newsletters cover stories from a single publication, which left us with a very small universe of individuals in our target population. All but one participant had many years of experience with curating newsletters, which likely allowed them to formulate (and articulate) a stable set of practices. However, we likely missed out on any challenges that journalists may face while incubating a fresh newsletter. All our participants curated news stories in English -- which has more volume and diversity of news compared to other languages. Future research should consider if curating news stories in other languages amounts to more or less challenges compared to English. In drawing a parallel between journalists' news curation and algorithmic aggregation, we have ignored any demand-side comparisons between the two. For instance, what benefits are perceived by the readers, and is there a clear preference for one over the other? While there is some evidence on the growing popularity of newsletters curated by journalists, prior work should look at a more detailed comparison of the two. 
\section{Findings}

\begin{figure*}
    \centering
    \includegraphics[width=0.8\textwidth]{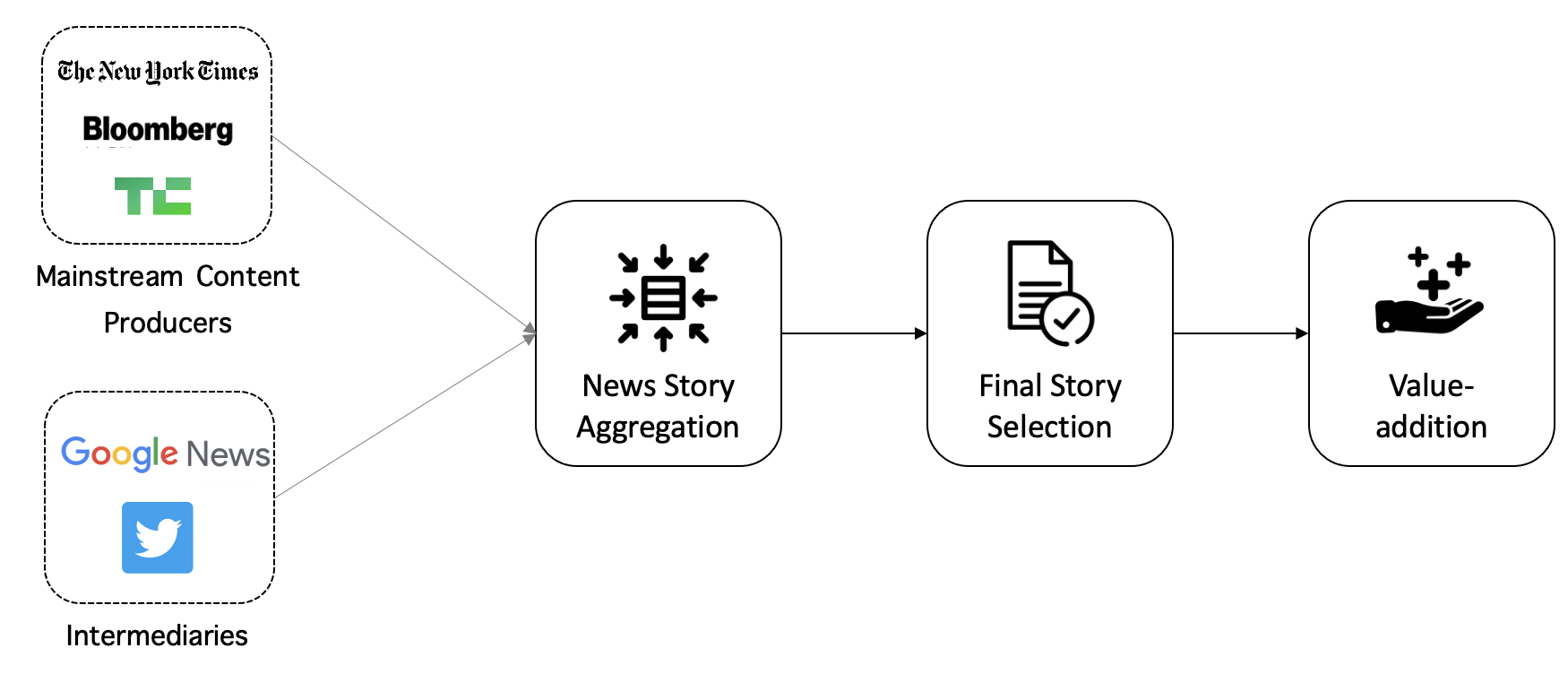}
    \caption{An overview of the news curation workflow}
    \label{fig:workflow}
\end{figure*}

We found that certain elements of the journalists' workflow for curating stories for newsletters were fairly consistent across both geographical regions and content domains. In Figure~\ref{fig:workflow}, we visualize their news curation workflow. We found that journalists use both mainstream original content producers (such as The New York Times, Bloomberg, etc.) and intermediary platforms (such as Twitter, Google News) to aggregate a large and diverse set of stories. When aggregating stories, journalists took into consideration the topic of the story as well as the perspectives covered in the story. When outlining their criteria for aggregation, journalists articulated the entire gamut of news values, such as relevancy, reference to elites, magnitude and significance, etc. \cite{harcup2001news, Harcup_ONeill_2017}. Stories that ascribed to one or more of these values were considered for aggregation. However, only stories that were evaluated as credible or authoritative were aggregated. 

The next step was typically an iterative process to shortlist a smaller set of stories that makes for a coherent compilation in the said newsletter edition. Curators also added value to the selected stories by providing additional context and/or highlighting important details, to excite their readers. These could be brief summaries, trivia, links to other stories, etc. While we visualize this as a flow, the steps for journalists differed---some preferred doing the value addition for specific stories before selecting others, while others made a full selection of stories before adding any commentary. 

While covering the findings from our interview, we first describe how curators navigate the news ecosystem of mainstream content producers and intermediaries (Section~\ref{sec:ecosystem}), followed by a description of the key factors that guide their workflow (Section~\ref{sec:factors}).

\subsection{Navigating the News Ecosystem}
\label{sec:ecosystem}

\subsubsection{Mainstream Content Producers}
Journalists who curate newsletters typically begin their process by aggregating their stories from a small set of high-recognition publishers such as The New York Times, The Wall Street Journal, etc. The reputation for quality and editorial control of these venues is an important element of what makes them attractive. These venues provide a ``safe starting point'' to begin the curation process.

While most journalists may agree on the reputation of a small set of such publications, each journalist had different preferred venues that they tended to engage repeatedly. These were domain specific -- thus business-oriented newsletters selected more from financial newspapers and journals, while journalists curating tech-oriented newsletters accordingly selected more of those sources.  

Journalists aggregated most of their stories by directly visiting the websites of these publishers, often doing that multiple times in a day to check for new stories. In this, journalists used their most preferred sources' websites as tickers -- using their updates as a means to keep up with unfolding stories. They used the form factor of the websites -- the placement of stories, size of fonts, expandable text etc. as signals for the prominence of stories. The website was the closest equivalent to the materiality of a physical newspaper, which journalists arguably couldn't really use since their job was to run through several sources quickly.

\begin{quote}
  ``There is just something that I like about visiting their own websites and homepages... it's the closest thing that I have to reading physical newspapers.''
\end{quote}

However, this quote also underlines the challenge with deep reading of individual sources. While automated approaches may lose some of the nuances of editorial signals in highlighting stories, the problem for a human editor is around picking and choosing which publishers they can dedicate significant attention to. In practical terms, such an approach may allow for a long-tailed editorial process in which a handful of sources get solid engagement while the overwhelming majority of sources get eyeballed only for the one or two stories that have potential for inclusion in a newsletter. This leads us to the next section, where we look at the role of intermediaries in the story selection process.




\subsubsection{Intermediaries}

Journalists used both social media platforms and news aggregators as part of their daily process of identifying news stories. The most mentioned aggregator was Google News, while the most mentioned social media platform was Twitter. \chiadd{While a few participants also reported using Google's search results in addition to Google News, most of them did not draw any explicit distinction between the two.}
Google News, which indexes stories from over 20,000 publishers \cite{searchenginejournalAvoidThese} is the leading aggregator in the world, and ranks among the top five news websites in the US \cite{statistaLeadingNews}, and the top three, in terms of trust \cite{statistaTrustNews}.
Twitter, on the other hand is the top social media channel for news consumption \cite{pewresearchNewsConsumption}. Both of these were critical professional resources that helped journalists ensure access to stories from a much wider set of publishers.



Most journalists seemed to have a basic understanding of the working of algorithms behind aggregators and social media platforms. For instance, journalists believed that the algorithms were impacted by popularity bias, and disproportionately represented certain news topics. Some journalists felt that this approach was too narrow as \textit{``how do human beings know what they want to know about until they see it''} (P3). Others took a more extreme stance and pointed out that popularity bias can adversely impact the quality of news aggregation as 
these algorithms are often \textit{``not prepared for what people will like''} (P2). 

This in turn meant that journalists were actively conscious about not reinforcing the algorithmic bias and avoided selecting top stories recommended by Google News, or stories that were trending on Twitter. In case of major events that were at \textit{``a crossover between trending and newsworthy'' } (P1) (e.g., the Ukraine-Russia war, the economic crisis in Sri Lanka), being aware of what was trending empowered journalists to take a counterprogramming \cite{bourreau2003mimicking} approach and curate stories that provided a new or an enhanced perspective. As P3 noted:

\begin{quote}
    ``When it came to today, we did the Sri Lanka lead because most of the other stories only reported about oh, they all went to the Presidential palace, they did this, they did that. But no one actually went beyond okay, what happens after this? If you Googled Sri Lanka protests, you will only find the clips that were getting the most clicks.''
\end{quote}

So while user engagement data such as link clicks or likes offer insight into popularity, journalists used these with caution. Unlike for a standard aggregator, for which the popularity of a story can be a single, deciding factor, journalists curating stories were often seeking novelty and serendipity - offering something that a typical algorithm may not.


Another strategy to offset algorithmic biases was the use of search and filter functionality provided by Google News. Journalists actively used these filters to find stories on particular topics which often led to a discovery of stories from atypical sources. As P4 noted: 

\begin{quote}
    ``One of the stories we did on Hotstar and Disney Plus, in the run-up to the IPL (Indian Premier League Cricket) media rights auction came from Variety... Variety is probably one of the least expected sources that you would expect a story on Hotstar-Disney to come from, but we took it from there because we felt it had value for our readers''
\end{quote}

\chiadd{More generally, journalists' strategies for overcoming algorithmic biases depended on the topics and the context they were curating for. For example, P1, who curated their newsletters weekly or biweekly, reported using time-based filters to offset the recency bias \cite{chakraborty2015can} inherent in algorithmic aggregators.}

A different challenge with social media was running into rabbit holes while seeking a good story for aggregation. Other than the reputation of the source, and the content of the story, journalist also had access to comments to a post about the story. This meant that the browsing process on social media was a lot more work than may be initially obvious. Some journalists dealt with this by using Twitters' ``Twitter Lists''\footnote{https://help.twitter.com/en/using-twitter/twitter-lists} functionality that enabled them to streamline their timeline by only seeing tweets from a restricted set of Twitter accounts that they trust. P8 highlighted the challenge of using social media:

\begin{quote}
    ``Being on social media does strange things to your notions of agency. I can go to social media with the idea that I'm going to find something. I'm gonna scroll through and see if there's anything I can grab for the newsletter, but I don't really know what I'm going to find once I get there.'' (P8)
\end{quote}

\subsection{Guiding Factors}
\label{sec:factors}
When outlining each of the step in their curation workflow -- aggregation, selection, and value-addition -- journalists pointed out different factors that guided their decisions and actions. These factors were fairly consistent across most participants and we summarize them as following -- (1) relying on their journalistic experience, (2) accounting for their readers' interests, and (3) acknowledging their subjective judgments.



\subsubsection{Journalistic Experience}


\begin{quote}
    ``Being journalists, we're not like academics or professionals, but we know something about many things.'' (P6)
\end{quote}

Journalists who curate newsletters typically reported having years of experience in researching and writing their own stories and reading stories written by their peers. Their experience equips them with an insider's perspective on the production and distribution of news in general, as well as the strengths and weaknesses of particular publishers and authors. A rule of thumb that most journalists adhered to was that the author of the story mattered more than where the story was published. This often meant that journalists had a preference for stories written by certain kinds of authors, for instance, domain experts and academics. P7 provided us with a relevant example:

\begin{quote}
    ``The Institute for the Study of War is a much better source for understanding developments about the Russian campaign in Ukraine, more than anything you're going to read in The Wall Street Journal or The Washington Post unless The Wall Street Journal story is written by Michael Gordon.''
\end{quote}

The advantage that journalists claimed was insider knowledge of exclusive content producers with demonstrated expertise. While algorithms may use popularity metrics to bring to focus highly engaged writers on a subject, familiarity with ``the expert to turn to'' for a specific issue is often lacking. During the interviews, our participants provided us with many such examples -- The Hakai Magazine for environment and coastal life (P2), Cabinet Magazine for arts and culture (P8), The Institute for the Study of War (P7, as pointed above), etc. Having this knowledge allowed journalists to proactively curate authoritative content rather than retroactively evaluating the authoritativeness of content reported by mainstream producers.  

For domains where this knowledge was lacking, journalists paid special attention to how the story was written. Reflecting on their own experience of writing stories, journalists pointed out various assessment signals for evaluating the information and arguments presented in the story. As a standard practice, journalists expected the information to be backed by multiple sources along with some description of how the information was sourced. Some journalists also looked for whether the arguments in a story were backed by quotes from independent experts or a citation to peer-reviewed research. As illustrated by P8:

\begin{quote}
    ``A classic example was a story about a guy who was in an officer-involved shooting, and the write-up is just from the police. Is the story really going to be what the police tell you? Or is it going to be more credible if you have eyewitnesses from the community who also say, yeah, the guy pulled the gun on the policeman versus if the eyewitnesses say, yeah, the cops shot him in the back.''
\end{quote}



By virtue of investing most of their time into reading and writing news stories, journalists are particularly equipped with the ability to find patterns across seemingly disparate stories. Even when curating stories for a newsletter, many journalists took on the extra work of finding these patterns and situating stories as part of larger narratives. Some journalists even reached out to their own trusted sources to gather additional knowledge that would help them contextualize the stories. This was often seen as an important value addition to the curation process, as many readers, especially young ones, are largely disconnected from everyday news and find it hard to follow \cite{Newman_Fletcher_Robertson_Eddy_Nielsen_2022}. P4 provided us with an example: 

\begin{quote}
    ``If it was only people visiting temples, then you could attribute it to many things. But then the astrology market is also going up, which means there are two different aspects of a person's solace seeking which are becoming businesses. This is what we call `connect the dots'. They are different stories but they are adjacent in their appeal.''
\end{quote}




\subsubsection{Readers' Interests}

\begin{quote}
    ``So as a mental model, you do the same thing as if you're selling chocolates, you have to get into the user's frame of mind.'' (P3)
\end{quote}

Unlike algorithms, manually curated newsletters are not personalized, and all readers get the same set of stories. Many newsletters, especially those with a geographical focus, invited a diverse set of readers. Readers may differ in terms of their demographics, as well as their news consumption habits. Some readers may consume news from multiple sources while others may simply rely on one newsletter. P8, who curates local news for an urban city, underlined their challenge:

\begin{quote}
    ``In terms of product design, thinking about who you are writing to, it could be somebody who has lived here for their whole life and is 70 years old, or it could be somebody who just moved here last week. So writing for both of those groups is very different but we try.''
\end{quote}

For some journalists, however, the interests of their readers were implicit in the objective of their newsletter. By making this objective explicit on the subscription page of their newsletter, journalists believed that they have curated an audience with a shared interest. Some journalists even constructed a detailed persona of who their readers might be. This persona further influenced the news topics they curated as part of their newsletter. P6, whose newsletter is focused on Business and Technology news for an Indian audience, provided a rich description: 

\begin{quote}
    ``Our core readership probably would be somewhere between 28 and 35, who is interested in finance, both personal as well as general finance. They are usually working in companies, probably mid-level to top-level managers. A lot of our readers are also startup entrepreneurs and founders and all that. They all like to know what's happening in their areas. And there would be gems [stories] which are probably hidden in some American newspaper, which we would think that they should know.''
\end{quote}


Nonetheless, the challenge of curating news for a diverse audience came up repeatedly. Journalists often tried to ensure that each newsletter edition contained an assortment of stories that catered to a diverse set of interests. For instance, a journalist curating business news would aggregate stories that speak to different industries, such as technology, retail, finance, etc. Journalists who cater to a global audience avoided selecting multiple stories from the same location. 



The affordances of their digital medium allow algorithmic aggregators and individual content publishers with their own websites, to collect detailed feedback on their readers' interests. Such feedback is often collected in terms of quantified metrics like number of links clicked, time spent on a page, etc. However, for journalists who circulate their newsletter via email, such feedback was often lacking. The only metric they could access was the number of readers who opened their email. 

In the absence of a more detailed feedback, many journalists paid special attention to any occasional comments they received from their readers. Some journalists took a more proactive approach in collecting this feedback by directly writing to their readers. P6 tried to take a hybrid approach by combining all they had: 

\begin{quote}
    ``We write to our readers very frequently. So depending on the analytics, which is provided by the platform, we can identify who are regular readers, who are sporadic readers, who don't read us at all. So I would write to each of these segments and ask them for feedback. Once or twice, we've done surveys in our newsletter itself, that what is it that you like, what is it that you don't like?''
\end{quote}



Instead of focusing on the specific interests of their readers, some journalists chose to emphasize on the reader-friendliness of their newsletter, i.e., by keeping it short and simple, and making it fun to read. In a media environment catered to decreasing attention spans of the users \cite{subramanian2018myth}, journalists found it particularly challenging to make news stories consumable. As a value-addition to their curation, some journalists extracted highlights from their stories, while others preferred to write their own crisp summaries. Such summaries could also include other elements that could not be gathered easily from a story - for instance how it aligned with contextual factors specific to an organization, or for instance how one article related to the rest of the selections for a newsletter.

As P9 noted, 


\begin{quote}
    ``I had to rewrite the blurb [for a story] a few times because when you have two sentences to explain something that is quite complex, you obviously have to pick and choose what elements you're going to highlight. [...] I've to write it in a way that is digestible, informative, not overly wordy, not overly boring, and something that will keep a person's attention throughout the email.''
\end{quote}

Not all curated stories, however, were complex or required additional summarization. As part of their curation process, journalists paid attention to aggregating a mix of stories that were light and fun to read. Many newsletters carried an additional section at the end where such stories were shared, perhaps as a way to counter the often highlighted negative effect of news on the readers' mood \cite{Newman_Fletcher_Robertson_Eddy_Nielsen_2022}. P3 told us about some of these stories: 

\begin{quote}
    ``We are not always about what is serious. You may want to know a cool story about some dinosaur that is selling for some stupid amount of money, some guy who actually set a painting on fire so that he could sell the NFT, or about a strange banana statue in Australia that was so ugly that the locals try to decapitate it.''
\end{quote}




\subsubsection{\chiadd{Context Dependence}\chiremove{ Subjective Judgments}}

\begin{quote}
    ``Some of this [news curation] is entirely taste-based and inexplicable. It almost feels like if I could eat 1000 dishes, how do I decide which ones are delicious?'' (P5)
\end{quote}

During our interviews, we found that all journalists agreed on a common set of news values and the traditional indicators of newsworthiness, such as the story should be important, impactful, or interesting in some way \cite{harcup2001news}. These indicators played an important role when journalists were aggregating stories for their newsletter. However, their judgments about what was considered important, impactful, or interesting were often influenced by their prior experiences\chiadd{, and the context they were curating for.} For instance, P3 considered it important to curate stories about powerful individuals, such as \textit{``if [Shinzo] Abe has been assassinated''} or \textit{``if [Elon] Musk has filed and said he's not gonna buy Twitter''}. These examples reflect a preference for news stories with a ``reference to elites'' -- an important news value covered in the literature \cite{harcup2001news}. On the other hand, P5, \chiadd{whose newsletter emphasizes a timeless criteria}, discounted the importance of this news value, and noted \textit{``we're not curating things that are only interesting because they're currently in the news''}. The comparison underlines a notion of relative importance that journalists often associated with different news values\chiadd{, depending on the curation context.}

Another point of difference appeared between journalists when considering the news value on magnitude and significance \cite{harcup2001news}. Some journalists preferred stories that were relevant to a broad audience (e.g., streaming services). Others preferred stories that were more impactful (e.g., gender rights), even if they attracted a potentially smaller audience.

When aggregating stories, some journalists also made their judgments based on how the story was framed. Framing a story often requires selective presentation of certain perspectives that are relevant to an issue \cite{entman1993framing}. Some journalists preferred stories that were written with a human interest frame. Stories written with a human interest frame often provide human examples and personal vignettes \cite{burscher2014teaching} to help the readers understand how the story impacts their day-to-day lives. Some journalists noted a preference for stories written with a conflict frame, i.e., stories that highlighted the disagreement(s) on an issue, often by referring to multiple sides or providing counter arguments \cite{burscher2014teaching}. More generally, the journalists' preference for a particular frame often depended on the topic of the story \chiadd{and the curation context}. P4, when reacting to a news story about Google portraying TikTok and Instagram as its competitor, noted:  

\begin{quote}
    ``Typically, the conventional way of thinking is that Google Search market is coming under threat. Right? But the angle that we bring in is also interesting because we are going in with the anti-trust angle.''
\end{quote}

\section{Discussion}

We interviewed 10 journalists from around the world who curate newsletters with stories selected from different publications. Our findings underlined the role of journalists' prior experience in the value they bring into the curation process, their own use of algorithms in finding stories for their newsletter, and their internalization of their readers' interests and the context they are curating for. In this section, we discuss how journalists' judgments contrast with algorithmic curation, and the potential of incorporating algorithms in journalists' existing workflows. We conclude with implications for future technologies to support the journalists' process of news curation.

\subsection{Human Subjectivity versus Algorithmic Judgment}




Journalists' prior experiences and interests were central to their news curation practices. While each of them put together content for different markets, these were common factors in what distinguished the human process of curation. For each journalist who curated stories, their individual sense of what drove their audiences was a factor they felt they selected stories on, but unlike an algorithm that has a clear sense of what drove those decisions, we do not see that in our interviews. This mirrors how traditional newsrooms operate in terms of editors' selection judgments \cite{Diakopoulos_Trielli_Lee_2021, carlson2018automating}. While traditional newsrooms may rely on advertising information or surveys of customers to make decisions, in general, this is an intangible element that \chiadd{still requires and} relies heavily on human judgment.\chiremove{ The journalist curating stories understands their process as ``algorithmic'' in that they learn from what stories seem to have more affective appeal.}

Journalists curating on a daily basis then have the impossible task of eyeballing the universe of daily \chiadd{published} content, and are consequently constrained by the attention they can put to a story or set of stories. \chiadd{In extention to the journalists' gatekeeping practices at the information gathering and publishing stage \cite{wallace2018modelling}, we find their curation practices are also influenced by an increased access to information sources (e.g. subject-specific experts), and the curation criteria impacted by their personal and journalistic context. For instance,} in our sample, we had some journalists who gave more weight to prominent personalities as a gauge for the appeal of stories, while others were quicker to mention timeliness or the style of coverage as their primary driver. At the same time, this subjectivity was not static, and journalists noted adjusting their drivers for shortlisting stories at different times. 

Algorithms, on the other hand, are designed to analyze large volumes of data. However, as we show in the background, algorithms encode their own logic of what is relevant knowledge \cite{gillespie2014relevance}, which is prone to various biases (e.g., popularity bias and homogeneity bias \cite{nikolov2019quantifying}) that are often reflected in their selection of news stories. As \citet{carlson2018automating} argues, it is important to acknowledge and critically analyze the ``judgment-rendering'' capacity of both humans and algorithms -- as these judgments are a reflection of their social, economical, and organizational context. For journalists, such judgment is often a result of learned orthodoxy \cite{carlson2018automating} as journalists build their own internalized models of social importance. Algorithms on the other hand are often governed by quantified estimates, such as increasing user engagement, which is closely tied to the economic incentives of the organization.

\chiadd{The key difference between algorithms and journalists boils down to the journalist's active influence on what is newsworthy for their readers. For instance, journalists paid special attention to the perspectives covered in a story while curating their newsletter. On the other hand, algorithmic aggregators, perhaps in a bid to remain objective, do not make any judgments based on the perspectives covered in a story. Instead, their judgments are based on indirect metrics, such as the popularity of a story. The journalists' curation process is very akin to the functioning of a traditional print editorial room where editors receive wire services from around the world and then select stories best suited for their publication.}\chiremove{ The key difference with the human system is the signal for affect. Every journalist reported reading every story to its entirety before adding it to a newsletter, and their personal reactions were the decisive factor in the choice to include a story.}


\subsection{Humans and Algorithms: A Hybrid Workflow}

\chiadd{Our findings add to the growing body of evidence in support of hybrid workflows that utilize algorithms to accomplish journalist tasks, such as curating the home page of a news app and sending push notifications \cite{bucher2017machines, diakopoulos20195}. Within the context of news curation from different sources, we find that} what humans could do in quality, algorithms did with scale. Algorithms were integrated into the workflow of every journalist we spoke with, and arguably, without algorithms, the work of newsletters would be significantly compromised. While there was some deep engagement (such as page by page reading) with single publications, all journalists relied on recommendations through aggregators and social media as feeders into their process. The advantage that journalists brought to the table was through the individual selection, and the value addition that made a set of stories a complete issue.

Algorithms by themselves have frequently been at the center of negative attention for creating filter bubbles and repeatedly reinforcing the same content. The recent Reuters Digital News report highlighted the problem of ``selective avoidance'' \citet{Newman_Fletcher_Robertson_Eddy_Nielsen_2022} of news, which was driven by the repetitiveness of news and its negative effect on mood. The report also found that the quantum of news is just too overwhelming for the average reader.

Our findings suggest that algorithms when combined with a professional journalist can have the exact opposite effect of limiting one's news consumption. When curating news stories, journalists repeatedly emphasized the importance of diversity and novelty in their news selection. One of the ways they enable this is precisely through letting algorithms point them in the right direction. Algorithms allow journalists to capture context on a story, to see how people react to it, and to find different perspectives on the same story using the search and filter functionality. The algorithm, thus, becomes a means for scalability. 




Fears of the further automation of news curation -- as seen through the automated feeds on Google News, as well as the more recent shift of Microsoft's news offerings entirely away from its previously human-edited front pages \cite{theguardianMicrosoftSacks} has added fuel to the notion of technology as a decimator of the journalistic profession. Developments in language technology also add fears about the summarization of stories from meta data, or the entire removal of editorial specialization from news curation. Added to this is the beleaguered state of print media around the world, with small local newspapers vanishing entirely \cite{nytimesOverNewspapers}, and journalism as a profession looking increasingly grim.

It is not surprising then that popular narratives on algorithms and journalism are presented in binaries by focusing on the 'removal' of the human from our news consumption \cite{qzAfterReplacing,cjrCenterNewsletter}. The existing work on algorithms overwhelmingly frames technology through bias, and its contribution to an overall ``worse mix of news'' as perceived by the readers \cite{shearer2019americans}. On the contrary, our work adds to a growing body of evidence that algorithms cannot completely substitute for journalists \cite{Diakopoulos_2019,carlson2018automating}. A deeper understanding of their practices further underlines the central role that technology can play in their news curation workflows.

With that in mind, it is important that technology design should be informed by the journalists' existing practices so that the assets of both journalists and technologies can be appropriately utilized. In the next subsection, we present key design recommendations for system designers looking to support journalists' news curation workflows.  


\subsection{Design Implications}

\chiadd{\textbf{Frame Identification:} We found that journalists went beyond the topic of a story to examine the frames and perspectives covered in an article to bring out nuances, and arguably biases in what they offered readers. A few journalists were able to put a finger on exactly what frames they sought such as a conflict frame, a human interest frame, a geopolitical frame etc. and while all did not articulate it in these exact terms, a sense of value addition was central to what a journalist felt their professional instinct brought to the table. Such frames differed based on the topic and context of a story. Selecting the appropriate frame is an important aspect of journalistic judgment, one that is largely ignored by algorithmic aggregators.  Yet, this is a place algorithms can be useful.
While there have been prior work \cite{nicholls2021computational, walter2019news, burscher2014teaching} exploring computational modeling of frames present in a news story, their application in supporting journalists' news curation workflow has not been explored. 
Our findings suggest that journalists will benefit from automatic identification of frames present in news stories to help make better selections. Building an accurate frame classifier is a complex problem, mainly because it requires a substantial amount of labeled data for training, and the labeling can only be performed by expert journalists.
If information about the frame(s) is available as metadata of a story, it will not only help journalists filter news stories written with their preferred frame(s) but will also help the newsreader sense the perspective of a story without reading the entire story.}

\chiadd{\textbf{Computing Novelty:} Our findings show that journalists focused on curating novel or under-the-radar stories for their newsletters, as a way to counter the popularity bias inherent in algorithmic aggregation. However, algorithmic aggregators only pick a handful of stories that are published every day. For instance, Apple News' algorithmic aggregator displays an average of 50 stories/day \cite{Bandy_Diakopoulos_2020}, while NYTimes alone publishes over 150 stories/day \cite{meyer2016many}. These numbers suggest that most stories remain under the radar, and consequently that discoverability is a problem for good writing as there is a likely bias against novelty in favor of what has standardizable appeal. This does represent an interesting research problem -- recent surveys on news aggregation and recommendation point that computing the novelty of a news story remains an open challenge \cite{raza2021news, karimi2018news}. In particular, novelty of a news story depends on the real-world event reported in that story as well as other reportings on that event. This is unlike other recommendation systems for music, movies, etc., where novelty of an item can be computed individually, for instance, by considering its popularity or whether a user has seen the item before \cite{raza2021news,karimi2018news}. Our findings support extending these directions. As an example, when journalists curated news on a major event (e.g., the economic crisis in Sri Lanka), they emphasised on finding stories that provided a novel perspective even though the topic is not novel. This suggests that the discovery of novel stories by a news recommendation algorithm could benefit from an initial clustering of stories based on the real-world event they are covering, and a subsequent understanding of perspectives covered in these stories. The frame identification module proposed above could be used for this task. Social media data will also be useful as stories widely shared on social media are no longer novel.    
Overall, the novelty of a story cannot be static or binary, as it will dynamically change with the inflow of newly published stories and news engagement patterns. Given the challenges inherent in computing the novelty of a story, existing algorithmic aggregators lack any support for finding novel stories. Other than journalists who curate their own newsletters, algorithmic aggregators will also benefit from an effective computation of novelty in news stories. For instance, algorithmic aggregators can provide a novelty-based ranking of stories in addition to chronological and popularity-based rankings that are currently available.}

\chiadd{\textbf{Bias Awareness and Mitigation:} A major criticism of algorithmic aggregation is the biases inherent in their selection of stories. Similarly, newsletters curated by journalists would likely exhibit their own biases. For instance, we found most of our journalists followed a long-tail curation process, wherein a majority of their curated stories belonged to a handful of content producers, whereas stories from all other content producers featured only sporadically. This can lead to homogeneity bias favoring a few publishers, as has been found in the case of algorithmic aggregators \cite{nikolov2019quantifying}. 
Future work can identify such biases in journalist-curated newsletters, similar to the bias-related work on algorithmic aggregators~\cite{nikolov2019quantifying,Bandy_Diakopoulos_2020}. Second, these biases can be presented to journalists in a consumable and effective way, such that they can both interrogate their own positions and act accordingly. Recent work on bias mitigation found that making journalists aware of their implicit biases can help reduce gender bias in their reporting \cite{kalra2021curbing}. Making journalists aware of their biases could be effective at reducing the biases in their newsletters. For instance, in case of homogeneity bias, each journalist can be provided with a monthly distribution of content publishers featured in their newsletters to nudge the journalist towards a more equitable representation of publishers. Similarly, providing journalists with a distribution of author demographics featured in their newsletters could reduce any gender bias implicit in their curation.}



\chiadd{\textbf{Tool Development:} All participants largely followed a manual process of curating stories for their newsletters. They searched for stories on individual content producer websites, algorithmic aggregators, and social media platforms.
Browsing manually, often multiple times a day, was found to be a time-consuming and cumbersome task. While prior work~\cite{Diakopoulos_Trielli_Lee_2021} has found success with designing integrated interfaces that gather information from multiple sources to help journalists discover newsworthy leads, our findings suggest that journalists curating their newsletters could also benefit from similar interfaces.
Apart from aggregating stories from multiple sources, such a tool can also help journalists by collecting social media engagement data for each news story. For instance, the tool can identify stories shared by a journalist's trusted sources on social media, and color code these stories to increase their visibility. Providing social media engagement data on individual stories will also enable journalists to select stories on topics or perspectives that have not yet received sufficient public attention. News stories can also be annotated with external links to (prior) similar stories or information on entities relevant to that story, to help journalists provide additional context on a story as part of their value addition. Other than bringing all information in one place, the tool can use artificial intelligence to mine additional insights from the story, such as the frames present in the story or its degree of novelty (as we noted in previous design insights). Although developing such a tool seems straightforward and a low-hanging fruit, we believe it is an important first step toward helping journalists curating their own newsletters. }

\section{Conclusion}


With algorithmic aggregators like Google News becoming a major source of online news consumption, questions have been raised about their functioning and the biases that are reflected in the selection of their stories. In this work, we looked at an alternate form of news aggregation -- one that is done by journalists curating newsletters. Through in-depth interviews, we investigated the various steps of their workflow, and found the crucial role of their expertise in selecting stories. We examined how human curation is not entirely human -- in that algorithms are a necessary part of their hybrid workflows and offered key insights on designing tools that can support the existing work of these experts. 

\begin{acks}
Thank you all our participants for their time and patience.
\end{acks}

\bibliographystyle{ACM-Reference-Format}
\bibliography{references}

\onecolumn

\appendix

\section{Codebook}
\label{appendix}

\begin{table}[h]

\begin{tabular}{llll}

\hline

\multicolumn{1}{|l}{\textbf{Theme} / Codes}    & \multicolumn{1}{l|}{\textbf{Count}} & \textbf{Theme} / Codes & \multicolumn{1}{l|}{\textbf{Count}} \\ \hline

\multicolumn{1}{|l}{\textbf{Navigating Content Producers (16.3\%)}} & \multicolumn{1}{l|}{\textbf{45}} & \textbf{Journalistic Experience (21.3\%)} & \multicolumn{1}{l|}{\textbf{59}} \\

\multicolumn{1}{|l}{Manual Process} & \multicolumn{1}{l|}{25} & Credibility & \multicolumn{1}{l|}{16} \\

\multicolumn{1}{|l}{Preferred publishers} & \multicolumn{1}{l|}{12} & Insider knowledge & \multicolumn{1}{l|}{12} \\

\multicolumn{1}{|l}{News cycle update} & \multicolumn{1}{l|}{5} & Preferred authors & \multicolumn{1}{l|}{10} \\

\multicolumn{1}{|l}{Wild card sources} & \multicolumn{1}{l|}{3} & Adding insights & \multicolumn{1}{l|}{10} \\
\cline{1-2}

\multicolumn{1}{|l}{\textbf{Navigating Intermediaries (21.3\%)}} & \multicolumn{1}{l|}{\textbf{59}} & Writing style & \multicolumn{1}{l|}{7} \\

\multicolumn{1}{|l}{Google news} & \multicolumn{1}{l|}{15} & Connecting stories & \multicolumn{1}{l|}{4} \\ \cline{3-4}

\multicolumn{1}{|l}{Diversity} & \multicolumn{1}{l|}{11} & \textbf{Readers' Interests (23.55\%)} & \multicolumn{1}{l|}{\textbf{65}} \\

\multicolumn{1}{|l}{Algorithmic bias}  & \multicolumn{1}{l|}{11} & Composition of stories & \multicolumn{1}{l|}{12} \\

\multicolumn{1}{|l}{Social media: twitter} & \multicolumn{1}{l|}{10} & Readability & \multicolumn{1}{l|}{11} \\

\multicolumn{1}{|l}{Using filters} & \multicolumn{1}{l|}{7} & Bringing novelty & \multicolumn{1}{l|}{11} \\

\multicolumn{1}{|l}{Popularity bias}  & \multicolumn{1}{l|}{5} & User feedback & \multicolumn{1}{l|}{7} \\
\cline{1-2}

\multicolumn{1}{|l}{\textbf{Context Dependence (17.3\%)}}  & \multicolumn{1}{l|}{\textbf{48}} & Fun and humor & \multicolumn{1}{l|}{7} \\

\multicolumn{1}{|l}{News values}  & \multicolumn{1}{l|}{16} & Diverse interests & \multicolumn{1}{l|}{7} \\

\multicolumn{1}{|l}{Preferred frames}  & \multicolumn{1}{l|}{9} & Background and summary & \multicolumn{1}{l|}{7} \\

\multicolumn{1}{|l}{Dynamic values}  & \multicolumn{1}{l|}{8} & Adding voice & \multicolumn{1}{l|}{3} \\

\multicolumn{1}{|l}{Curation niche} & \multicolumn{1}{l|}{8} &  & \multicolumn{1}{l|}{} \\

\multicolumn{1}{|l}{Subjectivity} & \multicolumn{1}{l|}{7} &  & \multicolumn{1}{l|}{} 

\\ \hline

\end{tabular}

\caption{Codebook from our analysis of interview transcripts. The codebook shows five themes (bold), 29 codes, prevalence (\%) for each theme, and the total count of each theme and code.}

\label{tab:codebook}

\end{table}

\end{document}